\documentclass[a4paper]{jpconf}
\usepackage{graphicx}
\begin{document}
\title{Clustering of Color sources and the Equation of State in Heavy Ion Collisions at RHIC and LHC Energies}

\author{R. P. Scharenberg}

\address{Department of Physics, Purdue University, West Lafayette, Indiana-47907, USA}

\ead{schrnbrg@purdue.edu}

\begin{abstract}
The initial temperature $T_{i}$, energy density $\varepsilon_{i}$, and formation time $\tau_{i}$ of the initial state of the QGP formed in the heavy ion collisions at RHIC and LHC energies are determined using the data driven Color String Percolation Model (CSPM). Multiparticle production by interacting strings stretched between projectile and target form a spanning cluster at the percolation threshold. 
The relativistic kinetic theory relation for $\eta/s$ is evaluated as a function of $\it T$ and the mean free path ($\lambda_{mfp}$) using data and CSPM. $\eta/s$($T_{i}$, $\lambda_{mfp}$) describes the transition from a strongly interacting QGP at $T/T_{c} \sim 1$ to a weakly coupled QGP at  $T/T_{c} \ge 6$.  We find that the reciprocal of $\eta/s$ is equal to the trace anomaly $\Delta = \varepsilon-3P/T^{4}$ which also describes the transition.
We couple this initial state of the QGP to a 1D Bjorken expansion to determine the sound velocity $c_{s}^{2}$ of the QGP for 0.85 $\le T/T_{c} \leq 3$. The bulk thermodynamic quantities and the equation of state are in excellent agreement with LQCD results.
\end{abstract}

\section{Clustering of Color Sources}

Multiparticle production is currently described in terms of color strings stretched between the projectile and the target, which decay into new strings and subsequently hadronize to produce observed hadrons. Color strings may be viewed as small areas in the transverse plane filled with color field created by colliding partons. With growing energy and size of the colliding system, the number of strings grows, and they start to overlap, forming clusters, in the transverse plane very much similar to disks in two dimensional percolation theory. At a certain critical density a macroscopic cluster appears that marks the percolation phase transition. This is the Color String Percolation Model (CSPM) \cite{pajares1,pajares2}. The interaction between strings occurs when they overlap and the general result, due to the SU(3) random summation of charges, is a reduction in multiplicity and an increase in the string tension hence increase in the average transverse momentum squared, $\langle p_{t}^{2} \rangle$. 
 We assume that a cluster of $\it n$ strings that occupies an area of $S_{n}$ behaves as a single color source with a higher color field $\vec{Q_{n}}$ corresponding to the vectorial sum of the color charges of each individual string $\vec{Q_{1}}$. The resulting color field covers the area of the cluster. As $\vec{Q_{n}} = \sum_{1}^{n}\vec{Q_{1}}$, and the individual string colors may be oriented in an arbitrary manner respective to each other , the average $\vec{Q_{1i}}\vec{Q_{1j}}$ is zero, and $\vec{Q_{n}^2} = n \vec{Q_{1}^2} $.

Knowing the color charge $\vec{Q_{n}}$ one can obtain the multiplicity $\mu$ and the mean transverse momentum squared $\langle p_{t}^{2} \rangle$ of the particles produced by a cluster of $\it n $ strings \cite{pajares2}
\begin{equation}
\mu_{n} = \sqrt {\frac {n S_{n}}{S_{1}}}\mu_{0};\hspace{5mm}                                    
\langle p_{t}^{2} \rangle = \sqrt {\frac {n S_{1}}{S_{n}}} {\langle p_{t}^{2} \rangle_{1}}
\end{equation} 
where $\mu_{0}$ and $\langle p_{t}^{2}\rangle_{1}$ are the mean multiplicity and $\langle p_{t}^{2} \rangle$ of particles produced from a single string with a transverse area $S_{1} = \pi r_{0}^2$. This implies a simple relation between the multiplicity and transverse momentum $\mu_{n}\langle p_{t}^{2}\rangle_{n}=n\mu_{0}\langle p_{t}^{2}\rangle_{1}$, which means conservation of the total transverse momentum produced. 

In the thermodynamic limit, one obtains an analytic expression \cite{pajares1,pajares2}
\begin{equation}
\langle \frac {n S_{1}}{S_{n}} \rangle = \frac {\xi}{1-e^{-\xi}}\equiv \frac {1}{F(\xi)^2}
\end{equation}
where $F\xi)$ is the color suppression factor and  
$\xi = \frac {N_{s} S_{1}}{S_{N}}$ is the percolation density parameter. 
Eq.(1) can be written as $\mu_{n}=n F(\xi)\mu_{0}$ and 
$\langle p_{t}^{2}\rangle_{n} ={\langle p_{t}^{2} \rangle_{1}}/F(\xi)$.  
The critical cluster which spans $S_{N}$, appears for $\xi_{c} \ge$ 1.2 \cite{satz}. 
It is worth noting that CSPM is a saturation model similar to the Color Glass Condensate (CGC),
 where $ {\langle p_{t}^{2} \rangle_{1}}/F(\xi)$ plays the same role as the saturation momentum scale $Q_{s}^{2}$ in the CGC model \cite{cgc,perx}. 

\section{Experimental Determination of the Color Suppression Factor  $F(\xi)$}

The suppression factor is determined by comparing the $\it pp$ and  A+A transverse momentum spectra. 
To evaluate the value of $\xi$ from data for Au+Au collisions, a parameterization of $\it pp$ events at 200 GeV  is used to compute the $p_{t}$ distribution 
\begin{equation}
dN_{c}/dp_{t}^{2} = a/(p_{0}+p_{t})^{\alpha}
\end{equation}
where a is the normalization factor.  $p_{0}$ and $\alpha$ are parameters used to fit the data. This parameterization  also can be used for nucleus-nucleus collisions to take into account the interactions of the strings \cite{pajares2}
\begin{equation}
dN_{c}/dp_{t}^{2} = \frac {a'}{{(p_{0}{\sqrt {F(\xi_{pp})/F(\xi)}}+p_{t})}^{\alpha}}
\end{equation}
The color suppression factor $F(\xi)$ is related to the percolation density parameter $\xi$.
\begin{equation}
F(\xi) = \sqrt \frac {1-e^{-\xi}}{\xi}
\end{equation}

\section{Energy Density $\varepsilon$}

The initial energy density $\varepsilon_{i}$ above $T_{c}$ is given by \cite{bjorken}
\begin{equation}
\varepsilon_{i}= \frac {3}{2}\frac { {\frac {dN_{c}}{dy}}\langle m_{t}\rangle}{S_{n} \tau_{pro}}
\end{equation}
To evaluate $\varepsilon_{i}$ we use the charged pion multiplicity $dN_{c}/{dy}$ at midrapidity and $S_{n}$ values from STAR for 0-10\% central Au-Au collisions with $\sqrt{s_{NN}}=$200 GeV \cite{levente}. We can calculate $ \langle p_{t}\rangle$ using the CSPM thermal distribution Eqs.(9) and (10). For $0.2 < p_{t} < 1.5 $, $\langle p_{t}\rangle = 0.394 \pm 0.003 $GeV, adding the extra energy required for the rest mass of pions at hadronization $\langle m_{t}\rangle = 0.42 \pm 0.003 $GeV. The error on $ \langle p_{t}\rangle$ is due to the error on $T_{i}$ \cite{eos}.

The dynamics of massless particle production has been studied in QE2 quantum electrodynamics.
QE2 can be scaled from electrodynamics to quantum chromodynamics using the ratio of the coupling constants \cite{wong}. The production time $\tau_{pro}$ for a boson (gluon) is \cite{swinger}  
\begin{equation}
\tau_{pro} = \frac {2.405\hbar}{\langle m_{t}\rangle}
\end{equation}
For Au-Au collisions at $\sqrt{s_{NN}}=$200 GeV Eqs. (6) and (7) gives  $\varepsilon_{i}$ = 2.27$\pm $0.16 GeV/$fm^{3}$ at $\xi$=2.88. In CSPM the  energy density $\varepsilon$ is proportional to $\xi$. 
From the measured value of  $\xi$ and $\varepsilon$ 
it is found  that $\varepsilon$ is proportional to $\xi$ for the range 
$1.2 < \xi < 10.56$, $\varepsilon_{i}= 0.788$ $\xi$ GeV/$fm^{3}$ \cite{levente,nucleo} as shown in Fig.1. 
\begin{figure}[thbp]
\centering        
\vspace*{-0.2cm}
\resizebox{0.55\textwidth}{!}{
  \includegraphics{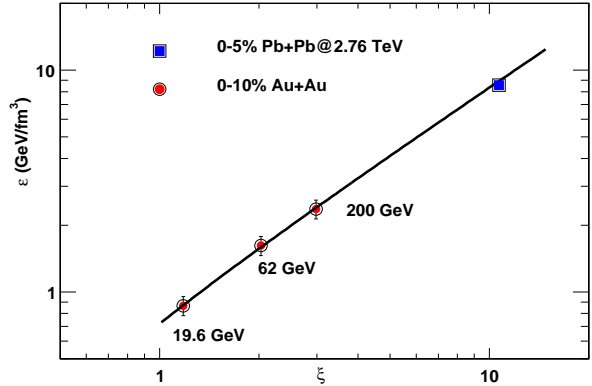}
}
\vspace*{-0.5cm}
\caption{Energy density $\epsilon$ as a function of the percolation density parameter $\xi$. The value for LHC energy is shown as blue square \cite{levente,nucleo}.} 
\end{figure}

\section{Determination of the Temperature}

The connection between the measured $\xi$ and the temperature $T(\xi)$ involves the Schwinger mechanism (SM) for particle production. 
The Schwinger distribution for massless particles is expressed in terms of $p_{t}^{2}$ \cite{swinger,wong}
\begin{equation}
dn/d{p_{t}^{2}} \sim e^{-\pi p_{t}^{2}/x^{2}}
\end{equation}
where the average value of the string tension is  $\langle x^{2} \rangle$. The tension of the macroscopic cluster fluctuates around its mean value because the chromo-electric field is not constant.
The origin of the string fluctuation is related to the stochastic picture of 
the QCD vacuum. Since the average value of the color field strength must 
vanish, it can not be constant but changes randomly from point to point \cite{bialas}. Such fluctuations lead to a Gaussian distribution of the string tension for the cluster, which transforms SM into the thermal distribution \cite{bialas}
\begin{equation}
dn/d{p_{t}^{2}} \sim e^{(-p_{t} \sqrt {\frac {2\pi}{\langle x^{2} \rangle}} )}
\end{equation}
with $\langle x^{2} \rangle$ = $\pi \langle p_{t}^{2} \rangle_{1}/F(\xi)$. 
The temperature is expressed as \cite{eos,pajares3}  
\begin{equation}
T(\xi) =  {\sqrt {\frac {\langle p_{t}^{2}\rangle_{1}}{ 2 F(\xi)}}}
\end{equation} 
 Recently, it has been suggested that fast thermalization in heavy ion collisions can occur through the existence of an event horizon caused by a rapid de-acceleration of the colliding nuclei \cite{khar2}. The thermalization in this case is due to the Hawking-Unruh effect \cite{hawk,unru}. In CSPM the strong color field inside the large cluster produces de-acceleration of the primary $q \bar q$ pair which can be seen as a thermal temperature by means of the Hawking-Unruh effect.    
The string percolation density parameter $\xi$ which characterizes the percolation clusters measures the initial temperature of the system. Since this cluster covers most of the interaction area, this temperature becomes a global temperature determined by the string density.
In this way at $\xi_{c}$ = 1.2 the connectivity percolation transition at $T(\xi_{c})$ models the thermal deconfinement transition.
  
We adopt the point of view that the experimentally determined universal chemical freeze-out temperature ($\it T_{f}$) is a good measure of the phase transition temperature, $T_{c}$. ${\langle p_{t}^{2}\rangle_{1}}$ is evaluated using Eq.(10) at $\xi_{c}$ = 1.2 with $\it T_{f}$ = 167.7 $\pm$ 2.6 MeV \cite{bec1}. This gives $ \sqrt {\langle {p_{t}^{2}} \rangle _{1}}$  =  207.2 $\pm$ 3.3 MeV. This calibrates the CSPM temperature scale. 

 Figure 2 shows a plot of $\varepsilon/T^{4}$  as a function of T/$T_{c}$. The lattice QCD results are from HotQCD Collaboration \cite{hotqcd}. It is observed that at LHC energy the CSPM results are in excellent agreement with the lattice QCD results. The lattice and CSPM results are available for T/$T_{c} < 2$. 
Table I gives the CSPM  values  $\xi$, $\it T$, $\varepsilon$ and $\eta/s$ at $T/T_{c}$ = 0.88, 1, 1.16 and 1.57. 

\begin{figure}[thbp]
\centering        
\vspace*{-0.2cm}
\includegraphics[width=0.55\textwidth,height=3.0in]{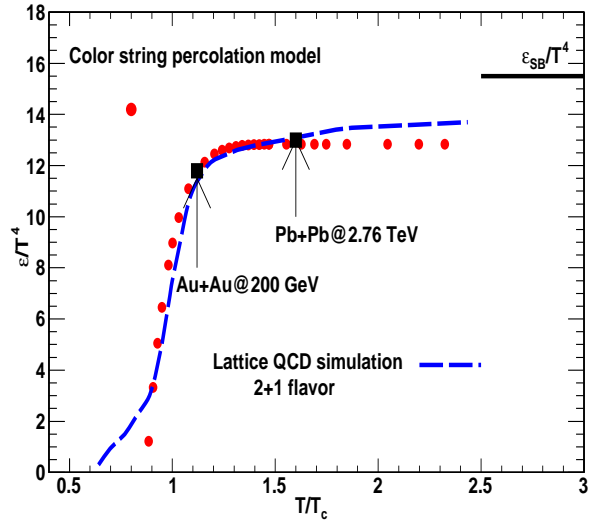}
\vspace*{0.0cm}
\caption{The energy density from CSPM  versus $T/T_{c}^{CSPM}$ (red circles) and Lattice QCD energy density vs  $T/T_{c}^{LQCD}$ (blue dash line) for 2+1 flavor and p4 action \cite{hotqcd}.} 
\end{figure}

\section{The Shear Viscosity to Entropy Density ratio $\eta/s$}
The relativistic kinetic theory relation for the shear viscosity over entropy density ratio, $\eta/s$ is given by \cite{gul1,gul2}
\begin{equation}
\frac {\eta}{s} \simeq \frac {T \lambda_{mfp}}{5}     
\end{equation}
where T is the temperature and $\lambda_{mfp}$ is the mean free path given by
\begin{equation}
\lambda_{mfp} \sim \frac {1}{(n\sigma_{tr})}
\end{equation}
$\it n $ is the number density of an ideal gas of quarks and gluons and $\sigma_{tr}$ the transport cross section for these constituents. 
\begin{table}
\caption{The measured percolation density parameter $\xi$, temperature $\it$T, energy density $\varepsilon$  and $\eta/s$ for the meson gas \cite{prakash};  the hadron to QGP transition;  Au+Au at 200 GeV  and Pb+Pb at 2.76 TeV . Au+Au is for 0-10$\%$ and Pb+Pb is for 0-5$\%$ central events.}
\vspace*{0.5cm}
\centering  
\setlength{\tabcolsep}{2pt}
\begin{tabular}{|c|c|c|c|c|}\hline
System & $\xi$ & T (MeV) & $\varepsilon (GeV/fm^3)$ & $\eta/s$ \\ \hline
 Meson Gas & 0.22 & 150.0 &- & 0.76 \\ \hline
 Hadron to QGP & 1.2 & 167.7 $\pm 2.6$ & 0.94$\pm 0.07$& 0.240$\pm 0.012$ \\ \hline
Au+Au & 2.88$\pm0.09$ & 193.6 $\pm3.0$ & 2.27$\pm 0.16$ & 0.204$\pm 0.020$ \\ \hline
Pb+Pb & 10.56$\pm1.05$ & 262.2$\pm13.0$ & 8.32$\pm 0.83$ &0.260$\pm 0.026$ \\ \hline 
\end{tabular}
\end{table} 

After the cluster is formed it behaves like a free gas of constituents. Eq. (10) can be applied to obtain the shear viscosity. In CSPM the number density is given by the effective number of sources per unit volume \cite{bautista} 
\begin{equation}
n = \frac {N_{sources}}{S_{N}L}
\end{equation}
 L is the longitudinal extension of the source, L = 1 $\it  fm $ \cite{pajares3}. The area occupied by the strings is related to $\xi$ through the relation $(1-e^{-\xi})S_{N}$. Thus the effective no. of sources is given by the total area occupied by the strings divided by the effective area of the string $S_{1}F(\xi) $. 
\begin{equation}
N_{sources} = \frac {(1-e^{-\xi}) S_{N}}{S_{1} F(\xi)} 
\end{equation}
 In general $N_{sources}$ is smaller than the number of single strings.  
The number density of sources from Eqs. (12) and (13) becomes
\begin{equation}
n = \frac {(1-e^{-\xi})}{S_{1}F(\xi) L}
\end{equation}
In CSPM the transport cross section $\sigma_{tr}$ is the transverse area of the effective string $S_{1}F(\xi)$. Thus $\sigma_{tr}$ is directly proportional to $F(\xi)$ and hence to $\frac {1}{T^{2}}$. The mean free path is given by
\begin{equation}
\lambda_{mfp} = {\frac {L}{(1-e^{-\xi})}} 
\end{equation}
For a large value of $\xi$ the $\lambda_{mfp}$ reaches a constant value. $\eta/s$ is obtained from $\xi$ and the temperature
\begin{equation}
\frac {\eta}{s} ={\frac {TL}{5(1-e^{-\xi})}} 
\end{equation}
 The behavior of $\eta/s$ is dominated by the fractional area covered by strings for $ \xi < \xi_{c}$. This is not surprising because $\eta/s$ is the ability to transport momenta at large distances and that has to do with the density of voids in the matter.
 
\begin{figure}[thbp]
\centering        
\vspace*{-0.2cm}
\resizebox{0.55\textwidth}{!}{
\includegraphics{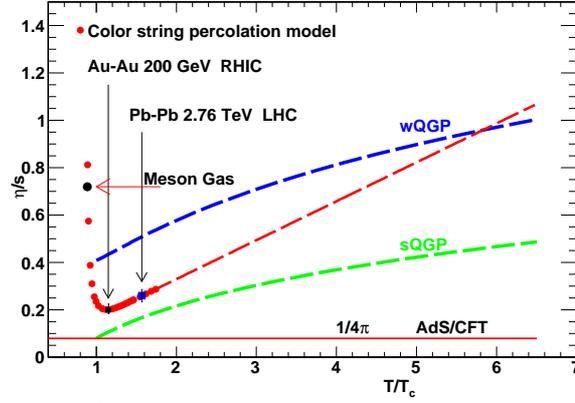}
}
\vspace*{-0.5cm}
\caption{$\eta/s$ as a function of T/$T_{c}$. Au+Au at 200 GeV for 0-10$\%$ centrality is shown as solid black square. wQGP and sQGP values are shown as dotted blue and green lines respectively \cite{gul1}. The red dotted line represents the extrapolation to higher temperatures from the CSPM. The hadron gas value for $\eta/s$ $\sim$ 0.7 is shown as solid black circle at T/$T_{c} \sim $0.88 \cite{prakash}.} 
\end{figure}       

 Figure 3 shows a plot of $\eta/s$ as a function of T/$T_{c}$. The estimated value of $\eta/s$ for Pb+Pb is shown in Fig. 3 at T/$T_{c}$ = 1.57.
The lower bound shown in Fig. 3 is given by AdS/CFT \cite{kss}. These results from STAR and ALICE data show that the $\eta/s$ value is 2.5 and 3.3 times the KSS bound \cite{kss}. 

The theoretical estimates of $\eta/s$ has been obtained as a function of T/$T_{c}$ for both the weakly (wQGP) and strongly (sQGP) coupled QCD plasma are shown in Fig. 3 \cite{gul1}. It is seen that at the RHIC top energy  $\eta/s$ is close to the sQGP. Even at the LHC energy it follows the trend of the sQGP. By extrapolating the $\eta/s$ CSPM values to higher temperatures it is clear that $\eta/s$ could approach the weak coupling limit near $T/T_{c}$ $\sim$ 6. 
 The CSPM  $\eta/s$ value for the hadron gas is in agreement with the calculated value using measured elastic cross sections for a gas of pions and kaons \cite{prakash}.
\begin{figure}[thbp]
\centering        
\vspace*{-0.2cm}
\includegraphics[width=0.55\textwidth,height=3.0in]{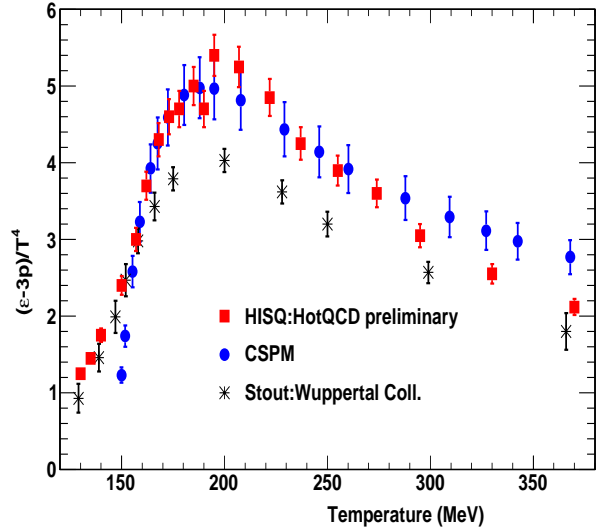}
\vspace*{0.0cm}
\caption{ The trace anomaly $\Delta =\varepsilon-3P/T^{4}$ vs temperature.} 
\end{figure}
\section{The trace Anomaly $\Delta$ and the reciprocal of $\eta/s$}
The $\Delta$ and $\eta/s$ both describe the transition from a strongly coupled QGP to a weakly coupled QGP. We find that the reciprocal of $\eta/s$ is in quantitative agreement with $\varepsilon-3P/T^{4}$ the trace anomaly over this wide temperature range. This result is shown in Fig.4. The minimum in $\eta/s$ =0.20 at $T/T_{c}$ 1.15 determines the peak of the interaction measure $\sim$ 5 in agreement with the recent HotQCD values\cite{lattice12}. 

\section{The Equation of State}
We use CSPM coupled to a 1D Bjorken expansion. The input parameters are the initial temperature T, the initial energy density $\varepsilon$, and the trace anomaly $\Delta$ are determined by data. The Bjorken 1D expansion gives the sound velocity   
\begin{eqnarray}
\frac {1}{T} \frac {dT}{d\tau} = - C_{s}^{2}/\tau  \\
\frac {dT}{d\tau} = \frac {dT}{d\varepsilon} \frac {d\varepsilon}{d\tau} \\
\frac {d\varepsilon}{d\tau} = -T s/\tau 
\end{eqnarray}
where $\varepsilon$ is the energy density, s the entropy density, $\tau$ the proper time, and $C_{s}$ the sound velocity. Since $s= \varepsilon + P/T$ and $P = (\varepsilon-\Delta T^{4})/3$ one gets
\begin{equation}
\frac {dT}{d\varepsilon} s = C_{s}^{2} 
\end{equation}
From above equations $C_{s}^{2}$ can be expressed in terms of $\xi$
\begin{equation}
 C_{s}^{2} = (-0.33)\left(\frac {\xi e^{-\xi}}{1- e^{-\xi}}-1\right)+\\
             0.0191(\Delta/3)\left(\frac {\xi e^{-\xi}}{({1- e^{-\xi}})^2}-\frac {1}{1-e^{-\xi}} \right)
\end{equation}

\begin{figure}[thbp]
\centering        
\vspace*{-0.2cm}
\includegraphics[width=0.55\textwidth,height=3.0in]{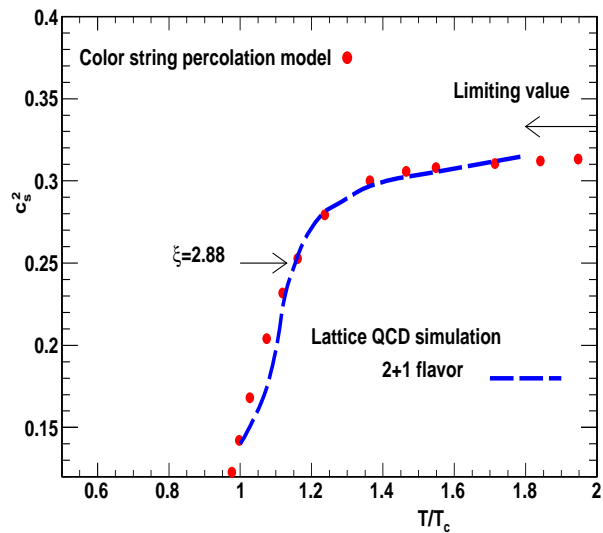}
\vspace*{0.0cm}
\caption{The speed of sound from CSPM  versus $T/T_{c}^{CSPM}$(red circles) and Lattice QCD-p4  speed of sound versus  $T/T_{c}^{LQCD}$(blue dash line)\cite{hotqcd}.} 
\label{perc9}
\end{figure}

\section{Conclusions and open questions}
Two central objectives in the experimental characterization of the QGP are the Equation Of State (EOS) and the shear viscosity to entropy ratio $\eta/s$. We have found that $s/\eta$ is equal to the trace anomaly $\Delta = \varepsilon - 3 P/T^{4}$. We determine the bulk thermodynamics, $\varepsilon/T^{4}$, $s/T^{3}$ and the EOS $c_{s}^{2}$ of the QGP as a function of $T/T_{c}$(CSPM). The results are  in excellent agreement with LQCD numerical simulations as a function of $T/T_{c}$(LQCD). CSPM predicts that the QGP will be formed in high multiplicity p-p collisions at LHC. The CSPM $\eta/s$ ratio can be used to test the quasi-particle hypothesis where the jet transport parameter is $\hat{q} = \frac {3}{2} T^{3} \frac {s}{\eta}$, which can be compared with $\hat{q}$ measured in jet quenching using the the ALICE EMCAL detector at LHC.
 
\section{Acknowledgments}
This research was supported by the Office of Nuclear Physics within the U.S. Department of Energy  Office of Science under Grant No. DE-FG02-88ER40412.

\end{document}